\documentstyle[aps,prl,epsf]{revtex}

\def\be{\begin{equation}}
\def\ee{\end{equation}}
\def\bea{\begin{eqnarray}}
\def\eea{\end{eqnarray}}
\def\sp{{\mathcal H}_2}
\def\ep{{\vec\epsilon}\ }
\newcommand{\la}{\langle}
\newcommand{\ra}{\rangle}
\begin{document}
\draft
\title{Three-party entanglement from positronium}
\author{A. Ac\'{\i}n, J. I. Latorre and P. Pascual}
\address{Departament d'Estructura i Constituents de la
Mat\`eria, Universitat de Barcelona, Diagonal 647, E-08028 Barcelona, Spain.\\
e-mail: acin@ecm.ub.es
}
\date{\today}
\maketitle
%%%%%%%%%%%% Abstract %%%%%%%%%%%%%%%%%%%%%%%%%%%
\begin{abstract}
The decay of ortho-positronium into three photons produces a
physical realization of a pure state with three-party
entanglement. Its quantum correlations are analyzed using recent
results on quantum information theory, looking for the final
state which has the maximal amount of GHZ-like correlations. This
state allows for a statistical dismissal of local realism
stronger than the one obtained using any entangled state of
two spin one-half particles.
\end{abstract}
\bigskip

\section{Introduction}
Entanglement or quantum correlations between many space-separated
subsystems has been recognized as one of the most intrinsic
properties of quantum mechanics and provides the basis for many
genuine applications of quantum information theory. It is, then,
quite natural to look for physical situations in which quantum
entangled states are obtained. Most of the theoretical and
experimental effort has so far been devoted to unveil physical
realizations of quantum states describing two quantum correlated
subsystems. The search for physical systems displaying clean
three-party entanglement is not simple. In this paper, we shall
analyze decays of particles as a natural scenario for fulfilling
such a goal. More precisely, we shall show that the decay of
ortho-positronium into three photons corresponds to a highly
entangled state. Let us now review what entanglement can be used
for and why it is interesting to look for quantum correlation
between more than two particles.

In 1935 Einstein, Podolsky and Rosen \cite{EPR}, starting from
three reasonable assumptions of locality, reality and
completeness that every physical theory must satisfy, argued that
quantum mechanics (QM) is an incomplete theory. They did not
question quantum mechanics predictions but rather quantum
mechanics interpretation \cite{GHZ}. Their argument was based on
some inconsistencies between  quantum mechanics and  their
local-realistic premises (LR) which appear for quantum states of
bipartite systems, $|\psi\ra\in{\mathcal H}_{d_1}\otimes{\mathcal
H}_{d_2}$. It was in 1964 when Bell \cite{Bell} showed that any
theory compatible with LR assumptions can not reproduce some of
the statistical predictions of QM, using a gedankenexperiment
proposed in \cite{BA} with two quantum correlated
spin-$\frac{1}{2}$ particles in the singlet state
 \be
|s\ra=\frac{1}{\sqrt 2}\left(|01\ra-|10\ra\right) .
 \ee
  In his
derivation, as it is well-known, quantum correlations or
entanglement have a crucial role. Actually,  the singlet state is
known to be the maximally entangled state between two particles.
The conflict between LR and QM arises since the latter violates
some experimentally verifiable inequalities, called Bell
inequalities, that any theory according to the local-realistic
assumptions ought to satisfy. It is then possible to design real
experiments testing QM against LR (for a detailed discussion see
\cite{CS}). Correlations of linear polarizations of pair of
photons were measured in 1982 showing strong agreement with
quantum mechanichs predictions and violating Bell inequalities
\cite{ADR}. Nowadays, Bell inequalities have been tested thoroughly in favor 
of QM \cite{exptest}.

More recently, it has been pointed out that some predictions for
quantum systems having quantum correlations between more than two
particles give a much stronger conflict between LR and QM than
any entangled state of two particles. The maximally entangled
state between three spin-$\frac{1}{2}$ particles, the so-called
GHZ (Greenberger, Horne and Zeilinger) state \cite{GHZor}
 \be
\label{GHZst} |{\rm GHZ}\ra=\frac{1}{\sqrt{2}}(|000\ra+|111\ra) ,
\ee
 shows some perfect correlations incompatible with any LR
model (see \cite{GHZ} and also \cite{Mer} for more details). It
is then of obvious relevance  to obtain these GHZ-like
correlations. Producing experimentally a GHZ state has turned out
to be a real challenge yet  a controlled instance  has been
produced in a quantum optics experiment \cite{BPDWZ}.

Entanglement is then important for our basic understanding of
quantum mechanics. Recent developments on quantum information
have furthermore shown that it is also a powerful resource for
quantum information applications. For instance, teleportation
\cite{BBCJPW} uses entanglement in order to obtain surprising
results which are impossible in a  classical context. A lot of
work has been performed trying to know how entanglement can be
quantified and manipulated. Our aim in this paper consists on
looking for GHZ-like correlations, which are truly three-party pure
state entanglement, in  the decay of
 ortho-positronium to three photons. The choice of this physical
system has been motivated mainly by several reasons. First, decay
of particles seems a very natural source of entangled particles.
Indeed, positronium decay to two photons was one of the physical
systems proposed long time ago as a source of two entangled
space-separated particles \cite{CHSH}. On a different line of
thought, some experiments for testing quantum mechanics have been
recently proposed using correlated neutral kaons coming from the
decay of a $\phi$-meson \cite{kaon}. In the case of
positronium, three entangled photons are obtained in the final
state, so it offers the opportunity of analyzing a quantum state
showing three-party correlations similar to other experiments
in quantum optics.

The structure of the paper goes as follows. We first review  the
quantum states emerging in both para- and  ortho-positronium
decays. Then, we focus on their entanglement properties and
proceed to a modern analysis of the three-photon decay state of
ortho-positronium. Using techniques developed in the context
of quantum information theory, we show that this state  allows in
principle for an experimental test of QM finer than the ones
based on the use of the singlet state. We have tried to make the paper
self-contained and easy to read for both particle physicists and
quantum information physicists. The first ones can found a
translation of some of the quantum information ideas to a
well-known situation, that is, the positronium decay to photons,
while the second ones can see an application of the very recent
techniques obtained for three-party entangled states, which allow
to design a QM vs LR test for a three-particle system in a
situation different from the GHZ state.

\section{Positronium decays}

\subsection{Positronium properties}

Let us start reminding some basic facts about positronium.
Positronium corresponds to a $e^+\,e^-$ bound state. These two
spin-$\frac{1}{2}$ particles can form a state with total spin
equal to zero,  para-positronium (p-Ps), or equal to one,
ortho-positronium (o-Ps). Depending on the value of its angular
momentum, it can decay to an even or an odd number of photons as
we shall see shortly.

Positronium binding energy comes from the Coulomb attraction
between the electron and the positron. In the non-relativistic
limit, its wave function is \cite{IZ}
 \be
\Psi(r)=\frac{1}{\sqrt{\pi a^3}}\,e^{-\frac{r}{a}}=
\int\frac{d^3p}{(2\pi)^{3/2}}\,e^{i\vec p\cdot\vec
r}\tilde\Psi(\vec p)=\int\frac{d^3p}{(2\pi)^{3/2}}\,e^{i\vec
p\cdot\vec r}\frac{\sqrt{8 a^3}}{\pi(1+a^2p^2)^2} , 
\ee 
where $a=\frac{2}{m\alpha}$, i.e. twice the Bohr radius of atomic
hydrogen, and $m$ is the electron mass. Note that the wave
function takes significant values only for three-momenta such that
$p\lesssim\frac{1}{a}\ll m$, which is consistent with the fact
that the system is essentially non-relativistic.

The parity and charge conjugation operators are equal to
 \be
U_P=(-1)^{L+1}\qquad U_C=(-1)^{L+S}, \ee
 where $L$ and $S$ are the
orbital and spin angular momentum. Positronium states  are then
classified according to these quantum numbers so that the ground
states are $^1S_0$, with $J^{PC}=0^{-+}$, for the p-Ps and
$^3S_1+^3D_1$, having $J^{PC}=1^{--}$, for the o-Ps.

 Positronium
is an unstable bound state that can decay to photons. Since a
$n$-photon state transforms as $U_C|n\gamma\ra=(-1)^n|n\gamma\ra$
under charge conjugation, which is an exact discrete symmetry for
any QED process such as the decay of positronium, we have that
the ground state of p-Ps (o-Ps) decays to an even (odd) number of
photons \cite{WR}. The analysis of the decay of positronium to
photons can be found in a standard QED textbook \cite{IZ}.
Para-positronium lifetime is about 0.125 ns, while for the case
of ortho-positronium the lifetime is equal to approximately 0.14
$\mu$s \cite{Czarn}.

The computation of positronium decays is greatly simplified due
to the following argument. The scale which controls the structure
of positronium is of the order of $\vert \vec p\vert\sim \alpha
m$. On the other hand, the scale for postrinomium annihilation is
of the order of $m$. Therefore, it is easy to prove that
positronium decays are only sensitive to the value of the wave
function at the origin. As a consequence, it is possible to
factor out the value of the wave function from the tree-level QED
final state computation \cite{IZ}.  A simple computation of
Feymann diagrams will be enough to write the precise structure of
momenta and polarizations which describe the positronium decays.
Furthermore, only tree-level amplitudes need to be computed since
higher corrections are suppressed by one power of $\alpha$. Let
us now proceed to analyze the decays of p-Ps and o-Ps in turn.

\subsection{Para-positronium decay}
Para-positronium ground state decays into two photons. Because of
the argument mentioned above, the determination of the two-photon
state coming from the p-Ps decay is simply given by the lowest
order Feynmann diagram of $e^+e^-\longrightarrow \gamma\gamma$.
Since   positronium is a non-relativistic particle to a very good
approximation, the three-momenta of $e^+$ and
 $e^-$ are taken equal to zero,
 and the corresponding spinors are replaced by a two-component
spin. This implies that the tree-level calculation of the
annihilation of p-Ps into two photons is equal to, up to
constants,
 \be \label{ampl}
 {\cal M}(e^+e^-\longrightarrow \gamma\gamma)\sim\chi^{c\dagger}_+M_2\chi_- ,
\ee
where (see \cite{IZ} for more details) $\chi_{\pm}$
 is the two-component spinor describing the fermions,
  $\chi^{c\dagger}\equiv\chi^Ti\sigma_2$, and $M_2$ gives
\be M_2=\sum_{perm}(\ep_1^*\times\ep_2^*)\cdot\hat k\,I_{2\times
2} \equiv A(\hat k_1,\lambda_1;\hat k_2,\lambda_2)\,I_{2\times 2}
, \ee where $\ep^*_i\equiv\ep^*(\hat k_i,\lambda_i)$ stands for
the circular polarization vector associated to the outgoing
photon $i$ and $I_{2\times 2}$ is the $2\times 2$ identity matrix. More
precisely, for a photon having the three-momentum vector
$\vec{k}=|\vec{k}|\hat{k}=|\vec{k}|
(\sin\theta\cos\phi,\sin\theta\sin\phi,\cos\theta)$, the
polarization vectors can be chosen
\begin{equation}
\label{epsilon}
 \ep(\hat k, \lambda)= -\frac{\lambda}{\sqrt 2}
 \left(\cos\theta \cos \phi- i \lambda
 \sin\phi,\cos\theta\sin\phi+i\lambda\cos\phi,-\sin \theta\right) ,
\end{equation}
where $\lambda=\pm1$ and they obey
\begin{eqnarray}
\label{epsprop}
 &&\hat k\cdot\ep(\hat k,\lambda)=0
 \qquad \hat k\times \ep(\hat k,\lambda)=-i \lambda
 \ep(\hat k,\lambda) \nonumber
 \\
 && \ep(\hat k_i,\lambda_i)\cdot\ep(\hat k_j,\lambda_j)
  =-{1\over 2}\left( 1-\lambda_i\lambda_j \hat k_i \cdot \hat k_j\right) .
\end{eqnarray}
From the expressions of the polarizaton vectors and the
three-momentum and energy conservation, it follows that the
scalar term $A$ is \be A(\hat k,\lambda_1;-\hat
k,\lambda_2)=-\frac{i}{2}(\lambda_1+\lambda_2), \ee and it
verifies \bea \label{scalar}
&& A(\hat k,+1;-\hat k,+1)=-A(\hat k,-1;-\hat k,-1) \nonumber\\
&& A(\hat k,+1;-\hat k,-1)=-A(\hat k,+1;-\hat k,-1)=0 .
\eea
The two fermions in the para-positonium ground state are in
the singlet state, $|S$$=$$0,S_z$$=$$0\ra=\frac{1}{\sqrt{2}}
\left(|\frac{1}{2},-\frac{1}{2}\ra-|-\frac{1}{2},\frac{1}{2}\ra\right)$,
 and then, using the previous relations for $A$ and (\ref{ampl}),
 the two-photon state resulting of the p-Ps desintegration is
\be
\label{p-Psdes}
|\psi_p\ra=\frac{1}{\sqrt{2}}\left(|++\ra-|--\ra\right) .
\ee
The two-photon state resulting from p-Ps decay is thus equivalent
 to a maximally entangled state of two spin-$\frac{1}{2}$ particles.
 This is a well-known result and was, actually, one of the physical system
 first proposed as a source of particles having the quantum
 correlations needed to test QM vs LR \cite{CHSH}.

\subsection{Ortho-positronium decay}
The ground state of ortho-positronium has $J^{PC}=1^{--}$ and,
due to the fact that charge conjugation is conserved, decays to
three photons. Repeating the treatment performed for the p-Ps
annihilation, the determination of the three-photon state
resulting from the o-Ps decay requires the simple calculation of
the tree-level Feynmann diagrams corresponding to
$e^+e^-\longrightarrow \gamma\gamma\gamma$. Its tree-level
computation gives, up to constants,
 \be \label{ampl3}
 {\cal M}(e^+e^-\longrightarrow \gamma\gamma\gamma)\sim\chi^{c\dagger}_+M_3\chi_- ,
\ee
and the $2\times 2$ matrix $M_3$ is equal to \cite{IZ}
\be
 M_3=\sum_{cyclic\,perm.}
 \left(\left(\ep^*_2 \cdot \ep^*_3-\vec\delta_2
  \cdot \vec\delta_3\right) \ep^*_1 +
  \left(\ep^*_2\cdot \vec\delta_3+
  \ep^*_3\cdot \vec\delta_2\right) \vec\delta_1
  \right) \cdot \vec\sigma ,
\ee
where
\begin{equation}
  \vec \delta_i=\vec k_i\times \ep^*_i .
\end{equation}
Using (\ref{epsprop}) we can rewrite $M_3$ in the following way
\begin{equation}
\label{amplvec}
 M_3\equiv \vec\sigma\cdot
 \vec V(\hat k_1,\lambda_1;\hat k_2,\lambda_2;\hat k_3,\lambda_3) ,
\end{equation}
where
\begin{eqnarray}
\label{vector}
 \vec V=
 &&\left( (\lambda_1-\lambda_2)(\lambda_2+\lambda_3)\ \ep^*(\hat
    k_1,\lambda_1) \left(\ep^*(\hat k_2,\lambda_2)\cdot
     \ep^*(\hat k_3,\lambda_3)\right)\right. \nonumber
 \\
 &&+(\lambda_2-\lambda_3)(\lambda_3+\lambda_1)\  \ep^*(\hat
    k_2,\lambda_2) \left(\ep^*(\hat k_3,\lambda_3)\cdot
     \ep^*(\hat k_1,\lambda_1)\right) \nonumber
 \\
 &&\left. +(\lambda_3-\lambda_1)(\lambda_1+\lambda_2)\  \ep^*(\hat
    k_3,\lambda_3) \left(\ep^*(\hat k_1,\lambda_1)\cdot
     \ep^*(\hat k_2,\lambda_2)\right)
 \right) .
\end{eqnarray}
Notice that the helicity  coefficient
$(\lambda_i-\lambda_j)(\lambda_j+\lambda_k)$ for the cyclic permutations
of $ijk$ explicitly enforces the vanishing of the $(+++)$ and $(---)$
polarizations,
\begin{equation}
 \vec V(\hat k_1,+;\hat k_2,+;\hat k_3,+)=
  \vec V(\hat k_1,-;\hat k_2,-;\hat k_3,-)= 0 .
\end{equation}
On the other hand, the rest of structures are different from zero
\begin{eqnarray}
\label{nzterms}
  && \vec V(\hat k_1,-;\hat k_2,+;\hat k_3,+)= 2\,
    \ep^*(\hat k_1,-) (1-\hat k_2\cdot\hat k_3) \nonumber
  \\
  && \vec V(\hat k_1,+;\hat k_2,-;\hat k_3,-)= 2\,
    \ep^*(\hat k_1,+) (1-\hat k_2\cdot\hat k_3) ,
\end{eqnarray}
and similar expressions for the other cyclic terms.

The original $e^+\,e^-$ in the ortho-positronium could
 be in any of the three triplet states. It can be shown,
  using (\ref{ampl3}) and (\ref{amplvec}), that when the initial
   positronium state is $|S$$=$$1,S_z$$=$$1\ra=|\frac{1}{2},\frac{1}{2}\ra$,
   the decay amplitude is proportional to $V_1+iV_2$,
while the same argument gives $-V_1+iV_2$ for
$|S$$=$$1,S_z$$=$$-1\ra=|-\frac{1}{2},-\frac{1}{2}\ra$ and
$-\sqrt{2} V_3$ for
$|S$$=$$1,S_z$$=$$0\ra=\frac{1}{\sqrt{2}}(|\frac{1}{2},-\frac{1}{2}
\ra+|-\frac{1}{2},\frac{1}{2}\ra)$. Now, considering the explicit
expressions of the polarization vectors (\ref{epsilon}), with
$\theta=\frac{\pi}{2}$ without loss of generality, and
(\ref{nzterms}), it is easy to see that the three-photon state
coming from the o-Ps decay is, up to normalization, \bea
\label{finst}
|\psi_0(\hat k_1,\hat k_2,\hat k_3)\ra = &&(1-\hat k_1\cdot \hat k_2)(|++-\ra+|--+\ra) \nonumber \\
+&&(1-\hat k_1\cdot \hat k_3)(|+-+\ra+|-+-\ra) \nonumber \\
+&&(1-\hat k_2\cdot \hat k_3)(|-++\ra+|+--\ra) ,
\eea
when the third component of the ortho-positronium spin, $S_z$, is equal to zero, and
\bea
\label{finst2}
|\psi_1(\hat k_1,\hat k_2,\hat k_3)\ra = &&(1-\hat k_1\cdot \hat k_2)(|++-\ra-|--+\ra) \nonumber \\
+&&(1-\hat k_1\cdot \hat k_3)(|+-+\ra-|-+-\ra) \nonumber \\
+&&(1-\hat k_2\cdot \hat k_3)(|-++\ra-|+--\ra) , \eea when
$S_z=\pm 1$.

 The final state of the o-Ps decay is, thus, an
entangled state of three photons, whose quantum correlations
depend on the angles among the momenta of the outgoing three
photons. For the rest of the paper we will consider the first
family of states ($S_z=0$) although equivalent conclusions are
valid for the second one. In the next sections we will analyze
the entanglement properties of the states $|\psi_0(\hat k_1,\hat
k_2,\hat k_3)\ra$, using some of the quantum information
techniques and comparing them to the well-known cases of the
singlet and GHZ state.

\section{Entanglement properties}
The quantum correlations of the three-photon entangled state
obtained from the o-Ps annihilation depend on the position of the
photon detectors, i.e. on the photon directions we are going to
measure. Our next aim will be to choose from the family of states
given by (\ref{finst}) the one that, in some sense, has the
maximum amount of GHZ-like correlations. In order to do this, we
first need to introduce some recent results on the study of
three-party entanglement.

 The set of states $|\psi_0(\hat
k_1,\hat k_2,\hat k_3)\ra$ form a six-parameter dependent family
in the Hilbert space $\sp\otimes\sp\otimes\sp$, so that each of
its components is equivalent to a state describing three
spin-$\frac{1}{2}$ particles or three qubits (a qubit, or quantum
bit, is the quantum version of the classical bit and corresponds
to a spin-$\frac{1}{2}$ particle). Two pure states belonging to a
generic composite system ${\mathcal H}_d^{\otimes N}$, i.e. $N$
parties each having a $d$-dimensional Hilbert space, are
equivalent as far as their entanglement properties go when they
can be transformed one into another by local unitary
transformations. This argument gives a lower bound for the
entanglement parameters a generic state $|\phi\ra\in {\mathcal
H}_2^{\otimes N}$ depends on. Since the number of real parameters
for describing it is $2^{N+1}$, and the action of an element of
the group of local unitary transformations $U(2)^{\otimes N}$ is
equivalent to the action of $U(1)\times SU(2)^{\otimes N}$, which
depends on $3N+1$ real parameters, the number of entanglement
parameters is bounded by $2^{N+1}-(3N+1)$. For our case this
counting of entanglement parameters gives six, since we have
$N=3$, and it can be proved that this is indeed the number of
nonlocal parameters describing a state in
$\sp\otimes\sp\otimes\sp$ \cite{Pop}.

The above arguments  imply that six independent quantities
invariant under the action of the group of local unitary
transformations will be enough, up to some discrete symmetry, to
describe the entanglement properties of any three-qubit pure
state. Given a generic state $|\phi\ra\in {\mathcal H}_2^{\otimes
3}$
 \be \label{genst} |\phi\ra=\sum_{i,j,k} t_{ijk}|ijk\ra\quad
i,j,k=1,2, \ee where $|i\ra,|j\ra,|k\ra$ are the elements of a
basis in each subsystem, A, B and C, the application of three
local unitary transformations $U^A$, $U^B$ and $U^C$ transforms
the coefficients $t_{ijk}$ into \be t_{ijk}'=\sum
U_{i\alpha}^AU_{j\beta}^BU_{k\gamma}^Ct_{\alpha\beta\gamma} . \ee
From this expression it is not difficult to build polynomial
combinations of the coefficient $t_{ijk}$ which are invariant
under local unitary transformations \cite{Pop,Sud}. These
quantities are good candidates for being an entanglement
parameter. For example, one of these invariants is \be \sum
t_{i_1j_1k_1}t_{i_1j_2k_2}^\ast
t_{i_2j_2k_2}t_{i_2j_1k_1}^\ast={\rm tr}(\rho_A^2), \ee where
$\rho_A={\rm tr}_{BC}(|\phi\ra\la\phi|)$ is the density matrix
describing the local quantum state of A (and the same happens for
B and C).
 In \cite{Sud} the six linearly independent polynomial invariants of
minor degree were found (a trivial one is the norm) and a
slightly modified version of these quantities was also proposed
in \cite{nos}. In the rest of the paper we will not consider the
norm, so the space of entanglement parameters of the normalized
states belonging to $\sp\otimes\sp\otimes\sp$ has dimension equal
to five.

A particularly relevant polynomial invariant is the so-called
tangle, $\tau$, introduced in \cite{CKW}. There is strong
evidence that somehow it is a measure of the amount of
``GHZ-ness'' of a state \cite{nos,CKW,DVC,BC}. It corresponds to
the modulus of the hyperdeterminant of the hypermatrix given by
the coefficients $t_{ijk}$ \cite{GKZ}, which from (\ref{genst})
corresponds to
 \be \label{tangle} \tau(|\phi\ra)=|{\rm
Hdet}(t_{ijk})|=\left|\sum \epsilon_{i_1i_2} \epsilon_{i_3i_4}
\epsilon_{j_1j_2} \epsilon_{j_3j_4} \epsilon_{k_1k_3}
\epsilon_{k_2k_4}t_{i_1j_1k_1}t_{i_2j_2k_2}t_{i_3j_3k_3}t_{i_4j_4k_4}\right|
,
\ee
where $\epsilon_{00}=\epsilon_{11}=0$ and $\epsilon_{01}=-\epsilon_{10}=1$.
This quantity can be shown to be symmetric under
permutation of the indices $i,j,k$.

\begin{figure}
\begin{center}
 \epsffile{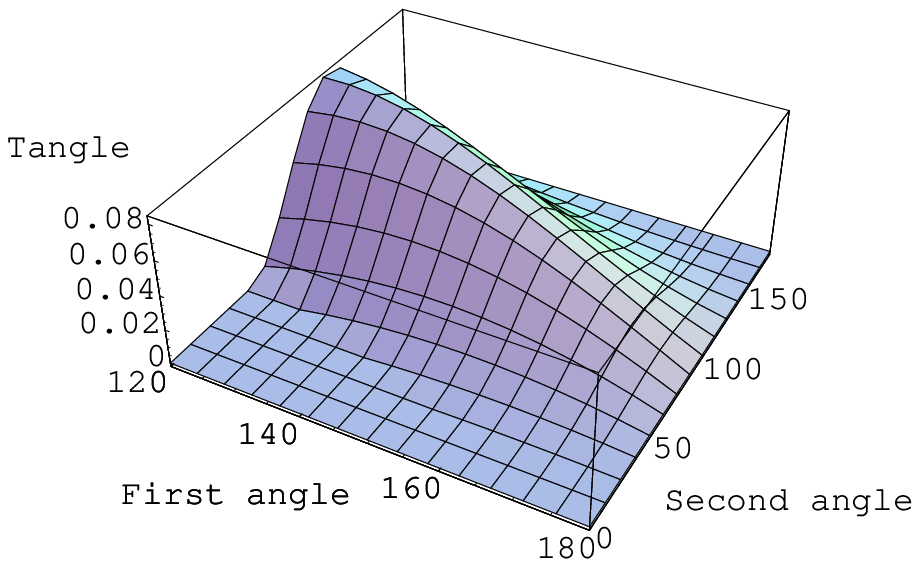}
\medskip
\caption{ Variation of the tangle with the position of the photon
detectors, that are represented by two angles (the third one has
to sum up to 360$^\circ$). We have taken $\tau=0$ when the
position of the detectors, i.e. the photon trajectories, are
incompatible with momentum conservation.} \label{tangledet}
\end{center}
\end{figure}

 Because of the interpretation
of the tangle as a measure of the GHZ-like correlations, we will
choose the position of the photon detectors, from the set of
states (\ref{finst}), the ones that are associated to a maximum
tangle. In the figure \ref{tangledet} it is shown the variation
of the tangle with the position of the detectors. It is not
difficult to see that the state of (\ref{finst}) with maximum
tangle corresponds to the case $\hat k_1\cdot \hat k_2=\hat
k_1\cdot \hat k_3=\hat k_2\cdot \hat k_3=-\frac{1}{2}$, i.e. the
most symmetric configuration, that we shall call ``Mercedes-star''
geometry. The normalized state obtained  from (\ref{finst}) for
this geometry is
 \be \label{state}
|\psi\ra=\frac{1}{\sqrt{6}}\left(|++-\ra+|--+\ra+|+-+\ra+|-+-\ra+|-++\ra+|+--\ra\right)
. \ee
 Note that the GHZ state has tangle equal to $\frac{1}{4}$,
while the value of the tangle of (\ref{state}) is lower,
\be
\label{stangle}
\tau(|\psi\ra)=\frac{1}{12} .
 \ee
It is arguable that a ``Mercedes-star'' geometry  was naturally
expected to produce a maximum tangle state. Indeed, GHZ-like
quantum correlations do not singularize any particular qubit.

 Let us also mention that the
state we have singled out has some nice properties from the point
of view of group theory. It does  correspond to the sum of two of
the elements of the coupled basis resulting from the tensor
product of three spin-$\frac{1}{2}$ particles,
$\frac{1}{2}\otimes\frac{1}{2}\otimes\frac{1}{2}$, \cite{Rai} \be
|\psi\ra=\frac{1}{\sqrt
2}\left(|\frac{3}{2},+\frac{1}{2}\ra+|\frac{3}{2},-\frac{1}{2}\ra\right)
, \ee where \bea
&&|\frac{3}{2},+\frac{1}{2}\ra=\frac{1}{\sqrt 3}\left(|++-\ra+|+-+\ra+|-++\ra\right) \nonumber\\
&&|\frac{3}{2},-\frac{1}{2}\ra=\frac{1}{\sqrt 3}\left(|--+\ra+|-+-\ra+|+--\ra\right) .
\eea
The quantum correlations of (\ref{state}) will be now analyzed.

\section{Useful decompositions}
In this section, the state (\ref{state}) will be rewritten in some
different forms that will help us to understand better its
nonlocal properties. First, let us mention that for any generic
three-qubit pure state and by performing change of local bases,
it is possible to make zero at least three of the coefficients
$t_{ijk}$ of (\ref{genst}) \cite{nos,HS}. A simple counting of
parameters shows that this is in fact the expected number of
zeros. This means that by a right choice of the local bases, any
state can be written with the minimum number of coefficients
$t_{ijk}$, i.e. we are left with all the non-local features of
the state, having removed all the ``superfluous'' information due
to local unitary tranformations. For the case of the state
(\ref{state}) it is easy to prove \cite{nos2} that it can be
expressed as 
\be 
\label{mindesc}
|\psi\ra=\frac{1}{2\sqrt{3}}\left(|001\ra+|010\ra+|100\ra\right)+\frac{\sqrt{3}}{2}|111\ra , 
\ee 
which is the minimum decomposition in terms of product
states built from local bases (four of the coefficients $t_{ijk}$
are made equal to zero).

An alternative decomposition, that will prove to be  fruitful for
the rest of the paper, consists of writing the state as a sum of
two product states. This decomposition is somewhat reminiscent of
the form of the GHZ state, which is a sum of just two product
states,  and is only possible when the tangle is different from
zero \cite{nos,DVC} as it happens for our state (see \ref{stangle}). The
state then can be written as
 \bea
\label{prdesc} |\psi\ra && =\frac{2}{3}\left(\left(\matrix{1 \cr
0}\right)\otimes\left(\matrix{1 \cr
0}\right)\otimes\left(\matrix{1 \cr
0}\right)+\left(\matrix{\frac{1}{2} \cr
\frac{\sqrt{3}}{2}}\right)\otimes\left(\matrix{\frac{1}{2} \cr
\frac{\sqrt{3}}{2}}\right)\otimes\left(\matrix{\frac{1}{2} \cr
\frac{\sqrt{3}}{2}}\right)\right)\nonumber\\
&&\equiv\alpha(|000\ra+|aaa\ra) , \eea
 where $|0\ra\equiv\left(\matrix{1 \cr 0}\right)$ and
$a\equiv\left(\matrix{\frac{1}{2} \cr \frac{\sqrt{3}}{2}}\right)$.
 We omit the details for the explicit computation of this
expression since they can be found in \cite{nos,DVC}. It is worth noticing
that o-Ps decay is hereby identified to belonging to an interesting type
of states already classified in quantum information theory \cite{DVC}.

 The above
decomposition allows for an alternative interpretation of the
initial state as an equally weighted sum of two symmetric product
states. Note that the Bloch vector, $\hat
n=(\sin\theta\cos\phi,\sin\theta\sin\phi,\cos\theta)$,
representing the first local spinor appearing in (\ref{prdesc})
is pointing to the $z$ axis, i.e. $\hat n_1=(0,0,1)$, while the
second is located in the $XZ$ plane with an angle of 120$^\circ$
with the $z$ axis, i.e.
$\hat{n}_2=(\frac{\sqrt{3}}{2},0,-\frac{1}{2})$. By performing a
new unitary transformation, (\ref{prdesc}) can be written as
 \be
\label{ghzdesc} |\psi\ra=\frac{2}{3}\left(\left(\matrix{c \cr
s}\right)\otimes\left(\matrix{c \cr
s}\right)\otimes\left(\matrix{c \cr s}\right)+\left(\matrix{s \cr
c}\right)\otimes\left(\matrix{s \cr
c}\right)\otimes\left(\matrix{s \cr c}\right)\right) ,
\ee
 where
$c=\cos15^\circ,s=\sin15^\circ$. Now, the two Bloch vectors are in
the $XZ$ plane, pointing to the $\theta=30^\circ$ and
$\theta=150^\circ$ directions. The GHZ state corresponds to the
particular case $c=1$ and $s=0$.

\section{Quantum mechanics vs local realism}
The quantum correlations present in some three-qubit pure states
show, as it was mentioned in the introduction, a much stronger
disagreement with the predictions of a local-realistic model than
any two-qubit entangled state. In fact, contrary to the case of
the singlet state, no LR model is able to reproduce all the
perfect correlations predicted for the maximally entangled state
of three qubits \cite{GHZ}. The state (\ref{state}) emerging from
o-Ps decay  is not a GHZ state, although it has been chosen the
one with the maximum tangle in order to maximize GHZ-like
correlations. In this section we will show how to use it for
testing quantum mechanics against local-realistic models, and
then we will compare its performance against  existing tests for
the maximally entangled states of two and three spin-$\frac{1}{2}$
particles. We start reviewing some of the consequences derived
from the arguments proposed in \cite{EPR}.

\subsection{QM vs LR conflict}
Given a generic quantum state of a composite system shared by $N$
parties, there should be an alternative LR theory which
reproduces all its statistical predictions. In this LR model, a
state denoted by $\lambda$ will be assigned to the system
specifying all its elements of physical reality. In particular,
the result of a measurement depending on a set of parameters
$\{n\}$ performed locally by one of the parties, say A, will be
specified by a function $a_\lambda(\{n\})$. The same will happen
for each of the space-separated parties and, since there is no
causal influence among them, the result measured on A can not
modify the measurement on B. For example, if the measurement is
of the Stern-Gerlach type, the parameters labeling the
measurement are given by a normalized vector $\hat n$ and
$a_\lambda(\hat n)\equiv a$ are the LR functions describing the
outcome.

The LR model can be very general provided that some conditions
must be satisfied. Consider a generic pure state belonging to
$\sp\otimes\sp\otimes\sp$ shared by three observers A, B and C,
which are able to perform Stern-Gerlach measurements in any
direction. Since the outcomes of a Stern-Gerlach measurement are
only $\pm1$, it is easy to check that for any pair of
measurements on each subsystem, described by the LR functions $a$
and $a'$, $b$ and $b'$, $c$ and $c'$, and for all their possible
values, it is always verified
 \be \label{LRconstr}
a'bc+ab'c+abc'-a'b'c'=\pm 2 . \ee
 It follows from this relation
that
 \be \label{mermineq} -2\leq\la a'bc+ab'c+abc'-a'b'c' \ra\leq
2. \ee
 This  constraint is known as Mermin inequality
\cite{Mermin} and has to be satisfied by any LR model describing
three space-separated systems.

Let us now take the GHZ state (\ref{GHZst}). It is quite simple to
see that if the observables $a$ and $a'$ are equal to $\sigma_y$
and $\sigma_x$ (the same for parties B and C), the value of
(\ref{mermineq}) is $-4$, so an experimental condition is found
that allows to test quantum mechanics against local realism. Note
that this is the maximal violation of inequality
(\ref{mermineq}). Moreover, the GHZ state  also satisfies that
$a'bc=ab'c=abc'=-a'b'c'=-1$ and no LR model is able to take into
account this perfect correlation result because of
(\ref{LRconstr}) \cite{GHZ}. This is a new feature that does not
appear for the case of a two maximally entangled state of two
spin-$\frac{1}{2}$ particles. In this sense it is often said that
a most dramatic contrast between QM and LR emerges for
entanglement between three subsystems.

Let us go back to the state given by the ortho-positronium decay
(\ref{state}). Our aim is to design an experimental situation
where a conflict between QM and LR appears, so we will look for
the observables that give a maximal violation of (\ref{mermineq}).
Such observables will extremize that expression.
Using the decomposition (\ref{ghzdesc}), the expectation value of
three local observables is \bea \label{expval}
\la abc \ra&=&\la\psi|(\hat n_a\cdot\vec\sigma)\otimes(\hat n_b\cdot\vec\sigma)\otimes(\hat n_c\cdot\vec\sigma)|\psi\ra \nonumber\\
&=&\frac{4}{9}\left(\prod_{i=a,b,c}(\tilde c\cos\theta_i+\tilde s\sin\theta_i
\cos\phi_i)+\prod_{i=a,b,c}(-\tilde c\cos\theta_i+\tilde s\sin\theta_i\cos
\phi_i)\right. \nonumber\\
&+&\left.\prod_{i=a,b,c}\sin\theta_i(c^2e^{-i\phi_i}+s^2e^{i\phi_i})+\prod_{i=a,b,c}\sin\theta_i(c^2e^{i\phi_i}+s^2e^{-i\phi_i})\right)
, \eea
where $\tilde c\equiv c^2-s^2$ and $\tilde s\equiv 2sc$.
 Because of the
symmetry of the state under permutation of parties, the
Stern-Gerlach directions are taken satisfying $\hat n_a=\hat
n_b=\hat n_c=(\sin\theta\cos\phi,\sin\theta\sin\phi,\cos\theta)$
and $\hat n_{a'}=\hat n_{b'}=\hat
n_{c'}=(\sin\theta'\cos\phi',\sin\theta'\sin\phi',\cos\theta')$.
 Substituting this expression in (\ref{mermineq}), we
get the explicit function $f(\theta,\phi,\theta',\phi')$ to be
extremized. For the case of the GHZ state described above, the
extreme values were obtained using two observables with
$\theta=\theta'=\frac{\pi}{2}$, i.e. in the $XY$ plane. Since
(\ref{ghzdesc}) is the GHZ-like decomposition of the initial
state, we take $\theta=\theta'=\frac{\pi}{2}$ and it is easy to
check that in this case $\frac{\partial
f}{\partial\theta}\big\vert_{\theta=\theta'=\frac{\pi}{2}}=\frac{\partial
f}{\partial\theta'}\big\vert_{\theta=\theta'=\frac{\pi}{2}}=0,\,\forall\phi,\phi'$.
Mantaining the parallelism with the GHZ case, it can be seen that
all the partial derivatives vanish when it is also imposed
$\phi=\frac{\pi}{2}$ and $\phi'=0$. In our case the calculation of
(\ref{mermineq}) gives $-3$, so a conflict between
local-realistic models and quantum mechanics again appears, and
then the three-photon state coming from the ortho-positronium
decay can be used, in principle, to test QM vs LR with the set of
observables given by the normalized vectors \be \label{test} \hat
n_a=\hat n_b=\hat n_c=(0,1,0)\qquad\hat n_{a'}=\hat n_{b'}=\hat
n_{c'}=(1,0,0) . \ee
 There is an alternative
set of angles $\phi$ and $\phi'$ that makes zero all the partial
derivatives of $f$: the combination of local observables
(\ref{mermineq}) is equal to $\approx-3.046$ for \be
\label{minalt}
\phi'=\arctan\left(-\frac{\sqrt{17+27\sqrt{41}}}{10}\right)\approx
126^\circ\quad\phi=\frac{1}{2}\arctan\left(\frac{2\sqrt{17+27\sqrt{41}}}{25}\right)\approx
24^\circ. \ee This second set of parameters will be seen to
produce in the end a weaker dismissal of LR.

 Our next step will be to carry over the comparison of this QM vs
LR test against the existent ones for the maximally entangled
states of three and two spin-$\frac{1}{2}$ particles, i.e. the GHZ
and singlet state. It is quite evident that the described test
should be worse than the obtained for the GHZ state. It is less
obvious how this new situation will compare with the singlet case.

\subsection{Comparison with the maximally entangled
 states of two and three spin-$\frac{1}{2}$ particles}

We will now  estimate the ``strength'' of the QM vs LR test
proposed above, being this ``strength'' measured by the number of
trials needed to rule out local-realism at a given confidence
level, as Peres did in \cite{Peres}. A reasoning anologous to the
one given in \cite{Peres} will be done here for the state
(\ref{state}) and the observables (\ref{test}).

 Imagine
a local-realistic physicist who does not believe in quantum
mechanics. He assigns prior subjective probabilities to the
validity of LR and QM, $p_r$ and $p_q$, expressing his personal
belief. Take for instance $\frac{p_r}{p_q}=100$. His LR theory is
not able to reproduce exactly all the QM statistical results of
some quantum states. Consider the expectation value of some
observable ${\mathcal O}$ with two outcomes $\pm 1$ such that
$\la{\mathcal O}\ra=E_q$ is predicted for some quantum state,
while LR gives $\la{\mathcal O}\ra=E_r\neq E_q$. Since the value
of the two possible outcomes are $\pm1$, the probablity of having
${\mathcal O}=+1$ is $q=\frac{1+E_q}{2}$ for QM and
$r=\frac{1+E_r}{2}$ for LR. An experimental test of the
observable ${\mathcal O}$ now is performed $n$ times yielding $m$
times the result $+1$. The prior probabilities $p_q$ and $p_r$
are modified according to the Bayes theorem and their ratio has
changed to \be
\frac{p_r'}{p_q'}=\frac{p_r}{p_q}\frac{p(m|_{LR})}{p(m|_{QM})} ,
\ee where \be p(m|_{LR})=\left(\matrix{n \cr
m}\right)r^m(1-r)^{n-m} , \ee is the LR probability of having $m$
times the outcome $+1$, and we have the same for $p(m|_{QM})$,
being $r$ replaced by $q$. Following Peres \cite{Peres}, the {\sl
confidence depressing factor} is defined \be
D\equiv\frac{p(m|_{QM})}{p(m|_{LR})}=\left(\frac{q}{r}\right)^m\left(\frac{1-q}{1-r}\right)^{n-m}
, \ee which accounts for the change in the ratio of the
probabilities of the two theories, i.e. it reflects how the LR
belief changes with the experimental results. Like in a game, our
aim is to destroy as fast as we can the LR faith of our friend by choosing an
adequate experimental situation. It can be said, for example,
that he will give up when, for example, $D=10^{4}$.
Since the world is quantum, $m=qn$, and the number of
experimental tests needed to obtain $D=10^{4}$ is equal to \be
\label{ntests}
n_D(q,r)\equiv\frac{4}{q\log_{10}\left(\frac{q}{r}\right)+(1-q)\log_{10}\left(\frac{1-q}{1-r}\right)}=\frac{4}{K(q,r)}
, \ee being $K(q,r)$ the information distance \cite{KU} between
the QM and LR binomial distribution for the outcome $+1$. The
more separate the two probability distributions are, measured in
terms of the information distance, the fewer the number of
experiments $n_D$ is.

 Let us come back to the three-party entangled state
coming from the ortho-positronium decay (\ref{state}) under the
local measurements described by (\ref{test}). As it has been
shown above, a contradiction with any LR model appears for the
combination of the observables given by the Mermin inequality. In
our case quantum mechanics gives the following predictions
 \be
\label{qmres} \la a'bc\ra=\la ab'c\ra=\la abc'\ra=-\frac{2}{3}
\qquad \la a'b'c'\ra=+1 , \ee
 and this implies that
$q_1=prob(a'bc=+1)=prob(ab'c=+1)=prob(abc'=+1)=\frac{1}{6}$ and
$q_2=prob(a'b'c'=+1)=1$. This is the QM data that our LR friend
has to reproduce as well as possible. Because of the symmetry of
the state he will assign the same probability $r_1$ to the events
$a'bc=+1$, $ab'c=+1$ and $abc'=+1$ and $r_2$ to $a'b'c'=+1$.
However, his model has to satisfy the constraint given by
(\ref{mermineq}), so the best he can do is to saturate the bound
and then 
\be 
\label{constraint}
3r_1=r_2\,\Longrightarrow\, 0\leq r_1\leq\frac{1}{3} . 
\ee 
Now, according to the probabilities $r_1$ and $r_2$ his LR
model predicts, we choose the experimental test that minimizes
(\ref{ntests}), i.e. we consider the event $a'bc=+1$
($a'b'c'=+1$) when $n_D(q_1,r_1)<n_D(q_2,r_2)$
($n_D(q_1,r_1)>n_D(q_2,r_2)$), and the experimental results will
destroy his LR belief after $n_D(q_1,r_1)$ ($n_D(q_2,r_2)$)
trials. The best value our LR friend can assign to $r_1$ is the
solution to 
\be 
\label{bestLR} 
n_D(q_1,r_1)=n_D(q_2,r_2) , 
\ee
with the constraint (\ref{constraint}), and this condition means that 
$r_1\approx0.315$ and $n_D\approx 161$ trials are needed
to have a depressing factor equal to $10^{4}$. Repeating the
same calculation for the observables giving by (\ref{minalt}),
the number of trials slightly increases, $n_D\approx 166$,
despite of the fact that the violation of the inequality is
greater than the obtained for (\ref{test}).

 In ref. \cite{Peres} the
same reasoning was applied to the maximally entangled state of
two and three spin-$\frac{1}{2}$ particles, showing that
$n_D\approx 200$ in the first case, and $n_D\approx 32$ for the
latter (see table \ref{summary}). Our result then implies that
the three-photon entangled state produced in the
ortho-positronium decay has, in some sense, more quantum
correlations than any entangled state of two spin-$\frac{1}{2}$
particles.

\subsection{Generalization of the results}
It is easy to generalize some of the results obtained for the
entangled state resulting from the o-Ps decay. As it has been
mentioned, this state can be understood as an equally weighted
sum of two symmetric product states, since it can be written as
(\ref{ghzdesc}). The Bloch vectors of the two local states
appearing in this decomposition form an angle of $120^\circ$. It
is clear that the conclusions seen above depend on the angle
between these vectors, i.e. with their degree of non-orthogonality.
The family of states to be analyzed can be parametrized in the
following way \be \label{famangle}
|\psi(\delta)\ra=\alpha_\delta\left(\left(\matrix{c_\delta \cr
s_\delta}\right)\otimes\left(\matrix{c_\delta \cr
s_\delta}\right)\otimes\left(\matrix{c_\delta \cr
s_\delta}\right)+\left(\matrix{s_\delta \cr
c_\delta}\right)\otimes\left(\matrix{s_\delta \cr
c_\delta}\right)\otimes\left(\matrix{s_\delta \cr
c_\delta}\right)\right) , \ee where $\delta$ is the angle between
the two local Bloch vectors,
$c_\delta\equiv\cos\left(\frac{\pi-\delta}{4}\right)$ and
$s_\delta\equiv\sin\left(\frac{\pi-\delta}{4}\right)$ and
$\alpha_\delta$ is a positive number given by the normalization of the state.
An alternative parametrization of this family is, using (\ref{mindesc}) and 
defining $\delta'\equiv\frac{\delta}{4}$,
\be
\label{famangle2}
|\psi(\delta)\ra=2\alpha_\delta\left(\sin^2\delta'\cos\delta'
\left(|001\ra+|010\ra+|100\ra\right)+
\cos^3\delta'|111\ra\right) . 
\ee 

\begin{figure}
\begin{center}
 \epsffile{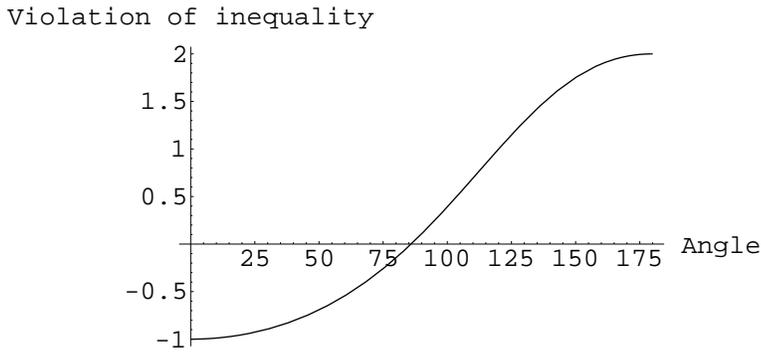}
\medskip
\caption{ Violation of the Mermin inequality (\ref{mermineq})
with the angle $\delta$ for the family of states
(\ref{famangle}). We have substracted 2 to the combination of the
expected values of (\ref{mermineq}), so a positive value means
that a conflict between QM and LR appears.} \label{mermangle}
\end{center}
\end{figure}

\begin{figure}
\begin{center}
 \epsffile{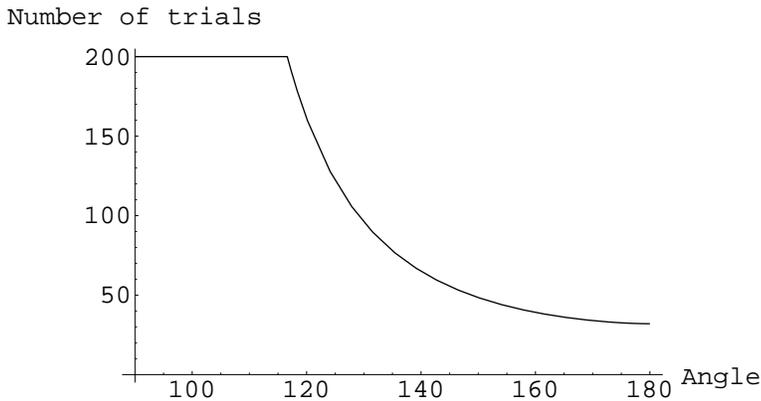}
\medskip
\caption{ Number of trials needed to rule out local realism as a
function of the angle $\delta$ for the family of states
(\ref{famangle}). Values greater than two hundred are not shown
since in these cases there always exists a two-qubit entangled
state which gives the same result, i.e. it has the same
``strength'' for ruling out local realism.} \label{ntrials}
\end{center}
\end{figure}

The expectation value of three local observables for this set of
states follows trivially from (\ref{expval}). Using this
expression it is easy to see that the combination of the
expectation values of (\ref{mermineq}) has all the partial
derivatives equal to zero for the set of observables given in
(\ref{test}) independently of $\delta$. For these observables,
the dependence of expression (\ref{mermineq}) with the degree of
orthogonality between the two product states is given in figure
\ref{mermangle}. There is no violation of the Mermin inequality
for the case in which $\delta\lesssim 85^\circ$. In this
situation one can always found a LR model able to reproduce the
QM statistical prediction given by (\ref{mermineq}) and the
observables (\ref{test}). We can now repeat all the steps made in
order to determine the number of trials needed to rule out local
realism as a function of the angle $\delta$. In figure
\ref{ntrials} we have summaryzed the results. We have shown only
the cases where the number of trials is minor than two hundred,
since this is the value obtained for the singlet. Note that the
case $\delta=120^\circ$, which corresponds to (\ref{state}), is
very close to the region where there is no improvement compared
to the maximally entangled state of two qubits.

All these results can be understood in the following way: the
smaller the angle between the two local states, $\delta$, the
higher the overlap of the state $|\psi(\delta)\ra$ with the
product state having each local Bloch vector pointing in the
direction of the $x$ axis, which corresponds to the state $|111\ra$ in 
(\ref{famangle2}). This means that the quantum state we
are handling is too close to a product state \cite{HS}, and thus, no
violation of the Mermin inequality can be observed.

\section{Concluding remarks}
In this work we have analyzed the three-particle quantum
correlations of a physical system given by the decay of the
ortho-positronium into a three-photon pure state. After obtaining
the state describing the polarization of the three photons
(\ref{state}), some of the recent techniques developed for the
study of three-party entanglement have been applied. The
particular case where the three photons emerge in a symmetric,
Mercedes-star-like configuration, corresponds to the state with
the maximum tangle. We have shown that this state allows a priori
for a QM vs LR test which is stronger than any of the existing
ones that use the singlet state.  In this sense, ortho-positronium
decays into a state which carries stronger quantum correlations
than any entangled state of two spin-$\frac{1}{2}$ particles.

Bose symmetrization has played a somewhat negative role in
reducing the amount GHZ-ness of the o-Ps decay state. Indeed, the
natural GHZ combination $\vert ++-\rangle +\vert --+\rangle$
emerging from the computation of Feynmann diagrams has been
symmetrized due to the absence of photon tagging to our state
$\vert ++-\rangle +\vert +-+\rangle+\vert -++\rangle +\vert
--+\rangle+\vert -+-\rangle +\vert +--\rangle$, inducing a loss
of tangle. The quantum optics realization of the GHZ state does
avoid symmetrization through a geometric tagging \cite{BPDWZ}. It is, thus,
reasonable to look for pure GHZ states in  decays to distinct
particles, so that  tagging would be carried by other quantum
numbers, as e.g. charge. It is, on the other hand, peculiar to
note that symmetrization in the $K^0\bar K^0$ system is
responsible for its entanglement ($\vert +-\rangle +\vert
-+\rangle$) \cite{kaon}.

Finally, let us briefly discuss the experimental requirements
needed for testing quantum mechanics as it has been described in
this paper. In order to do this, the circular polarizations of the
three photons resulting from an ortho-positronium decay have to be
measured. The positions of the three detectors are given by the
``Mercedes-star'' geometry and their clicks have to detect the
coincidence of the three photons. The energy of these photons is
of the order of 1 Mev. Polarization analyzers with a good
efficiency would allow us to acquire statistical data showing
quantum correlations which would violate the Mermin inequality
discussed above. Unfortunately, as far as we know, no such
analizers exist for this range of energies. A possible way-out
might be to use Compton scattering to measure the photon
polarizations \cite{AHSHZ}. However, Compton effect just gives a
statistical pattern depending on the photon and electron
polarizations which is not a direct measurement of the
polarizations. Further work is needed to modify our analysis of
QM vs LR to accommodate for such indirect measurements.

\bigskip
\section*{Acknowledgments}
We acknowledge J. Bernabeu for suggesting positronium as a source
of three entangled particles and reading carefully the paper. We
also thank  A. Czarnecki, D. W. Gidley, M. A. Skalsey and V. L.
Telegdi for comments about the measurement of the photon
polarizations in the ortho-positronium decay. We acknowledge
financial support by CICYT project AEN 98-0431, CIRIT project
1998SGR-00026 and CEC project IST-1999-11053, A. A. by a grant
from MEC (AP98). Financial support from the ESF is also
acknowledged. This work was concluded during the 2000 session of the Benasque
Center for Science, Spain.

\newpage

\begin{table}
\begin{tabular}{||c|lc||}
  ~~~~~~~~~~~~~~~~~~~~~~~~~~~~~~~~~~~~~~ State ~~~~~~~~~~~~~~~~~~~~~~~~~~~~~~~~~~~~~~ &
   Number of trials & \\
\hline
  GHZ & $\approx 32$ & \\
  Positronium state (\ref{state}) & $\approx 161$ & \\
  Singlet & $\approx 200$ & \\
\hline
\end{tabular}
\medskip
\caption{Comparison of the strength of the QM vs LR test which
can be performed for the maximally entangled states of two and
three spin-$\frac{1}{2}$ particles and for the three-photon
entangled state reulting from the ortho-positronium annihilation.}
\label{summary}
\end{table}


\begin{references}
\bibitem{EPR}
A. Einstein, B. Podolsky and N. Rosen, Phys. Rev. {\bf 47} (1935), 777.
\bibitem{GHZ}
D. M. Greenberger, M. A. Horne, A. Shimony and A. Zeilinger, Am. J. Phys. {\bf 58} (1990), 1131.
\bibitem{Bell}
J. S. Bell, Physics {\bf 1} (1964), 195.
\bibitem{BA}
D. Bohm and Y. Aharonov, Phys. Rev. {\bf 108} (1957), 1070.
\bibitem{CS}
J. F. Clauser and A. Shimony, Rep. Prog. Phys. {\bf 41} (1978), 1881.
\bibitem{ADR}
A. Aspect, J. Dalibard and G. Roger, Phys. Rev. Lett. {\bf 49} (1982), 1804.
\bibitem{exptest}
W. Tittel, J. Brendel, H. Zbinden and N. Gisin, Phys. Rev. Lett. {\bf 81} (1998), 3563; G. Weihs, T. Jennewein, C. Simon, H. Weinfurter and A. Zeilinger, Phys. Rev. Lett. {\bf 81} (1998), 5039.
\bibitem{GHZor}
D. M. Greenberger, M. A. Horne and A. Zeilinger, ``Going beyond Bell's theorem'', in {\sl Bell's Theorem, Quantum Theory and Conceptions of the Universe}, edited by M. Kafatos (Kluwer Academic, Dordrecht, The Netherlands, 1989), pp. 73-76.
\bibitem{Mer}
N. D. Mermin, Am. J. Phys. {\bf 58} (1990), 731.
\bibitem{BPDWZ}
D. Bouwmeester, J. Pan, M. Daniell, H. Weinfurter and A. Zeilinger, Phys. Rev. Lett. {\bf 82} (1999), 1345, quant-ph/9810035.
\bibitem{BBCJPW}
C. H. Bennett, G. Brassard, C. Cr\'epeau, R. Josza, A. Peres and W. K. Wootters, Phys. Rev. Lett. {\bf 70} (1993), 1895.
\bibitem{CHSH}
J. F. Clauser, M. A. Horne, A. Shimony and R. A. Holt, Phys. Rev. Lett. {\bf 23} (1969), 880.
\bibitem{kaon}
See for instance A. Di Domenico, Nucl. Phys. B {\bf 450} (1995), 293;  B. Ancochea, A. Bramon and M. Nowakowski, Phys.Rev. D {\bf 60} (1999), 094008, hep-ph/9811404;  F. Benatti and R. Floreanini, Eur. Phys. J. C {\bf 13} (2000), 267, hep-ph/9912348.
\bibitem{IZ}
C. Itzykson and J. Zuber, {\sl Quantum field theory}, McGraw-Hill.
\bibitem{WR}
L. Wolfenstein and D. G. Ravenhall, Phys. Rev. {\bf 88} (1952), 279.
\bibitem{Czarn}
Andrzej Czarnecki, Acta Phys. Polon. B {\bf30} (1999) 3837, hep-ph/9911455.
\bibitem{Pop}
N. Linden and S. Popescu, Fortsch. Phys. {\bf 46} (1998), 567,
quant-ph/9711016.
\bibitem{Sud}
A. Sudbery, ``On local invariants of pure three-qubit states'',
quant-ph/0001116.
\bibitem{nos}
A. Ac\' \i n, A. Andrianov, L. Costa, E. Jan\'e, J.I. Latorre and R. Tarrach,``Schmidt decomposition and classification of three-quantum-bit states'', quant-ph/0003050, to appear in Phys. Rev. Lett..
\bibitem{CKW}
V. Coffman, J. Kundu, W. K. Wootters, Phys. Rev. A {\bf 61} (2000), 052306,
quant-ph/9907047.
\bibitem{DVC}
W. D\"ur, G. Vidal and J. I. Cirac, ``Three qubits can be entangled in two inequivalent ways'', quant-ph/0005115.
\bibitem{BC}
T. A. Brun and O. Cohen, ``Parametrization and distillability of three-qubit entanglement'', quant-ph/0005124.
\bibitem{GKZ}
I. M. Gelfand, M. M. Kapranov and A. V. Zelevinsky, ``Discriminants,
resultants and multidimensional determinants'', Birkh\"auser Boston 1994.
Its explicit form is: ${\rm Hdet}(t_{ijk})=t_{000}^2t_{111}^2+t_{001}^2t_{110}^2+t_{010}^2t_{101}^2+t_{100}^2t_{011}^2$ $-2(t_{000}t_{111}t_{011}t_{100}+t_{000}t_{111}t_{101}t_{010}+t_{000}t_{111}t_{110}t_{001}$ $+t_{011}t_{100}t_{101}t_{010}+t_{011}t_{100}t_{110}t_{001}+t_{101}t_{010}t_{110}t_{001})$ $+4(t_{000}t_{110}t_{101}t_{011}+t_{111}t_{001}t_{010}t_{100})$.
\bibitem{Rai}
S. Rai and J. Rai, ``Group-theoretical Structure of the Entangled States of N Identical Particles'', quant-ph/0006107.
\bibitem{HS}
A. Higuchi and A. Sudbery, ``How entangled can two couples get?'', quant-ph/0005013; H. A. Carteret, A. Higuchi and A. Sudbery, ``Multipartite generalisation of the Schmidt decomposition'', quant-ph/0006125.
\bibitem{nos2}
A. Ac\'\i n, A. Andrianov, E. Jan\'e, J. I. Latorre and R. Tarrach, in preparation.
\bibitem{Mermin}
N. D. Mermin, Phys. Rev. Lett. {\bf 65} (1990), 1838.
\bibitem{Peres}
A. Peres, ``Bayesian analysis of Bell inequalities'', quant-ph/9905084.
\bibitem{KU} S. Kullback, {\sl Information theory and statistics}, Wiley, New York (1959).
\bibitem{AHSHZ}
B. K. Arbic, S. Hatamian, M. Skalsey, J. Van House and W. Zheng, Phys. Rev. A {\bf 37} (1988), 3189.

\end{references}
\end{document}